%% file: main.tex
\documentclass[authoryear]{article}

\usepackage{sectsty}
\usepackage{graphicx}
\usepackage{amsmath}
\usepackage{amssymb}
\usepackage{multicol}
\usepackage{changes}
\usepackage{bbm}
\usepackage{float}
\usepackage{booktabs}
\usepackage{subcaption}
\usepackage{rotating}
\usepackage[font={small,it}]{caption}
\usepackage{multirow}
\usepackage{comment}
\usepackage{tabularx}
\usepackage{natbib}
\usepackage{authblk}
\makeatletter
\newcommand*\bigcdot{\mathpalette\bigcdot@{.5}}
\newcommand*\bigcdot@[2]{\mathbin{\vcenter{\hbox{\scalebox{#2}{$\m@th#1\bullet$}}}}}

\excludecomment{mycomment}
\makeatother

	\title{A spatial modeling framework for monitoring surveys with different sampling protocols with a case study for bird abundance in mid-Scandinavia }
	\author{}

\author[1]{Jorge Sicacha-Parada \thanks{jorge.sicacha@ntnu.no}}
\author[2]{Diego Pavon-Jordan \thanks{diego.pavon-jordan@nina.no}}
\author[1]{Ingelin Steinsland \thanks{ingelin.steinsland@ntnu.no}}
\author[2]{Roel May \thanks{roel.may@nina.no}}
\author[2]{B\aa rd Stokke \thanks{bard.stokke@nina.no}}
\author[3]{Ingar Jostein \O ien \thanks{ingar@birdlife.no}}

\affil[1]{Department of Mathematical Sciences. Norwegian University of Science and Technology (NTNU). Trondheim, Norway }
\affil[2]{Department of Terrestrial Ecology, Norwegian Institute for Nature Research (NINA), P.O. Box 5685 Torgarden, N-7485. Trondheim, Norway}
\affil[3]{Norwegian Ornithological Society-BirdLife Norway, Sandgata 30 B, NO-7012. Trondheim, Norway}

\date{}

\begin{document}

\input{titlepage.tex}

\input{introduction.tex}

\input{data.tex}

\input{models.tex}

\input{simulation.tex}

\input{results.tex}

\input{discussion}

\input{acknowledgements.tex}

		\section*{Bibliography}
		\begingroup
		\renewcommand{\section}[2]{}%
		\bibliography{TOVERefs}
		\endgroup

\end{document}

%% file: titlepage.tex
\begin{titlepage}
	
\maketitle

\begin{abstract}
	
Quantifying the total number of individuals (abundance) of species is the basis for spatial ecology and biodiversity conservation. Abundance data are mostly collected through professional surveys as part of monitoring programs, often at a national level. These surveys rarely follow exactly the same sampling protocol in different countries, which represents a challenge for producing biogeographical abundance maps based on the transboundary information available covering more than one country. Moreover, not all species are properly covered by a single monitoring scheme, and countries typically collect abundance data for target species through different monitoring schemes.\\ \\
We present a new methodology to model total abundance by merging count data information from surveys with different sampling protocols. The proposed methods are used for data from national breeding bird monitoring programs in Norway and Sweden. Each census collects abundance data following two different sampling protocols in each country, i.e. these protocols provides data from four different sampling processes. The modeling framework assumes a common Gaussian Random Field shared by both the observed and true abundance  with either a linear or a relaxed linear association between them. The models account for particularities of each sampling protocol by including  terms that affect each observation process, i.e. accounting for differences in observation units and detectability. Bayesian inference is performed using the Integrated Nested Laplace Approximation (INLA) and the Stochastic Partial Differential Equation (SPDE) approach for spatial modeling. We also present the results of a simulation study based on the empirical census data from mid-Scandinavia to assess the performance of the models under model misspecification. Finally, maps of the 
total expected abundance of birds in our study region in mid-Scandinavia are presented with uncertainty estimates.\\ \\
We found that the framework allows for consistent integration of data from surveys with different sampling protocols. Further, the simulation study showed that models with a relaxed linear specification are less sensitive to misspecification, compared to the model that assumes linear association between counts. Relaxed linear specifications of total bird abundance in mid-Scandinavia improved both goodness-of-fit and the predictive performance of the models.

\end{abstract}
	
\end{titlepage}

%% file: introduction.tex
	\section{Introduction}
	
Understanding why organisms are where they are and what drives changes in their abundances is one of the main pillars of spatial ecology \citep{brodie2020} and is critical to propose effective measures to preserve biodiversity. In this regard, species distribution models (SDMs) have typically been used to gain a better understanding of species-habitat relationships \citep{brodie2020,Bradter2021} and to guide conservation practitioners and policy makers \citep{Araujo2019}. Previous SDMs using abundance data have revealed higher predictive performance in comparison with those using occurrence data \citep{Howard2014, Johnston2015}.
Yet, the majority of SDMs published to date used presence/absence (i.e. occurrence) data \citep{Araujo2019,yu2020}, rather than abundance data (count of individuals), especially in large-scale studies \citep{Milleretal2019}. This limits our ability to robustly infer, for example, regions with high density of individuals \citep{Johnston2015}, which is of paramount importance in conservation \citep{massimino2017}. For example, estimating abundance hotspots can inform and help authorities to select sites that may qualify to be included in the network of protected areas. Indeed, one of the main criteria to identify important areas for conservation under the European Union's Bird Directive (i.e. Special Protection Areas; SPA) is that a site accommodates regularly 1\% of the total biogeographical population of a species of conservation concern or more than 20,000 individuals of wetland birds (EU’s Birds Directive, 2009/147/EC 2009). Moreover, this Directive states that \textit{"The measures to be taken must apply to the various factors which may affect the numbers of birds, namely the repercussions of man’s activities and in particular the destruction and pollution of their habitats[...]"}. Abundance data can also be useful to detect and predict areas where human-wildlife conflicts may arise (e.g. \citet{may2020}), informing the corresponding authorities that infrastructure and further human development such as siting of powerlines and wind farms must be planned carefully (e.g. \citet{DeLucas2008} and \citet{may2020}).
Information about abundance is ultimately requested by national (e.g. Directorates, Environmental Agencies) and international (e.g. European Commission) authorities as basis to propose biodiversity conservation policies at different scales. This information should be based on all available count data.
\\ \\
Most countries have monitoring programs following national law and as signatories of international biodiversity conservation Directives and Conventions. These different national monitoring schemes may cover the same taxon (e.g. most countries have a national monitoring scheme for breeding birds) but can differ in the species recorded (different set of species may occur in the different countries and at different densities) and, most importantly, they usually follow different sampling protocols, which makes the information obtained by the different schemes not directly comparable. 
Furthermore, 
not all species are well represented in the data gathered within a single `general' protocol. For this reason, many countries have, for example, additional targeted monitoring schemes that complement the information for species that are considered poorly represented in the more general monitoring scheme. For example, colonial birds such as herons in Greece, raptors and waterbirds in Finland, nocturnal birds in Spain; see also \citet{buckland2017}.\\ \\
National common bird monitoring schemes and those targeting particular (groups of) species, provide together the largest datasets known on species abundance in time and space. However, at the (sub)national level, these datasets have mainly been used independently \citep{Kalas2010,Bevanger2014,Kery2009,Soykan2016} and multi-country studies have mostly analysed these data either independently for each country to later draw common conclusions from the country-specific estimates \citep{Lehikoinen2019} or combining the raw data with limited account for sampling differences (e.g. total abundance of waders; \citet{lindstrom2019}). Thus, overlooking the potential of integrating such a large amount of standardized data seems like an under usage of the effort and resources spent in collecting these data, especially when the taxa included in such monitoring schemes are very diverse, allowing not only to carry out species-specific analyses but also, potentially, community-level studies.\\ \\
This study was motivated by the need for estimates of the total abundance of birds in mid-Scandinavia based on high quality (i.e. standardized surveys)
localized data on bird abundances
from the common breeding bird monitoring programs in Norway (TOV-E) and Sweden (BBS). An estimate of the total abundance of birds can be used as an input for models that inform on the risk of infrastructure development (e.g. new powerlines and wind farms) for birds. The TOV-E and the BBS both provide standardized count data, but they differ in their sampling protocols. Both countries collect observations in point counts and transect surveys. In Norway, the main point counts (all species recorded) are complemented with line transects (only a subset of `rare' species also included in point counts are recorded -  see further details in Section 2). However, in Sweden, the  line transects and the point counts can be regarded as two different censuses (i.e. all species are counted in both census methods). These differences present the challenge of integrating the four sources of spatial information (points and transects in both Norway and Sweden) with different sampling protocols into one estimate for the spatial distribution of bird abundance for the entire region of interest \citep{brodie2020,Gruss2019}. 
\\ \\
The scarcity of studies applying large-scale abundance SDMs is likely related to (i) the generally lower availability of abundance data compared to occurrence data for most species
(\citet{Milleretal2019},
\citet{buckland2017} and references therein), and (ii) statistical and computational challenges of modelling abundance data.
Great methodological advancements to overcome some of these problems have been developed in the past decade, especially for integrating different data types, see \citet{Milleretal2019} and references therein. 
Most of these efforts have focused on 
enabling the use of
casually-collected (non-standardized) presence-only data to increase spatial coverage and data points of certain species (see also \citet{buckland2017}).
The possibility of improving 
SDMs by integrating abundance (count) data collected under different standardized monitoring schemes is most often neglected.  Thus, in addition to the integration of data from different countries, merging data from  different schemes (from one or several countries) can thus improve the estimates of abundance obtained from all available count data.
\\ \\
Given the existing gap in methodology for proper integration of standardized 
count data, we here propose a generic modelling framework that integrates standardized count data from various monitoring schemes (i.e. designed surveys) with different sampling protocols.
The models can ultimately  
produce one single estimate of
abundance (total abundance of birds in our case study) 
and its uncertainty based in data from different sampling protocols. In addition, it also gives interpretable estimates of the ecological parameters driving this abundance. Our methodology, thus, analyzes these data in a unique, single framework to produce models that account for different sampling processes, and describe and predict the spatial distribution of abundance.
\\ \\
Spatial modeling of multiple data sources has been approached for example in the context of coregionalization models \citep{Banerjee2015,blangiardo2015spatial,Krainski2019} and recently reviewed in \citet{Milleretal2019}. These are multivariate models for measurements that vary jointly over a region and have been defined through a hierarchical structure and fitted using Markov Chain Monte Carlo (MCMC) techniques \citep{Banerjee2015}. For the family of Spatial Latent Gaussian Models \citep{Rue2005}, the INLA-SPDE approach \citep{Rue2009,Lindgren2011} and its easy implementation in the INLA library of R have emerged as a faster alternative to jointly model multiple sources of information. Such method has been applied to multivariate models related with, for example, air pollution data \citep{Cameletti2019} and hydrology \citep{roksvag2020}. The proposed framework framework assumes
the existence of a latent process, underlying all the observed abundances, that represents the true expected abundances. 
The true expected abundance vary in space through spatial covariates as well as a spatial random effect.
Given the true expected abundance we assume that the observed abundances follow Poisson distributions. 
For each observation process a linear relation between the expected counts and the true expected abundances is assumed.
Further, we assume the existence of a common spatial random effect that drives the observed counts (cf. \citet{Milleretal2019}) for all the observation processes.   Given that the linear assumption may not depict the true relationship between the expected counts and the true expected abundances, we also propose models that allow deviations from this assumption. The proposed models are suitable doing computational fast inference using the INLA-SPDE approach, which approximates the posterior densities of parameters and hyperparameters. 
\\ \\
To the best of our knowledge, methodologies for jointly modeling spatial abundance using data from multi-country standardized biodiversity monitoring programs with different sampling protocols have not been published before. By properly integrating data from different monitoring schemes our method can be part of solving some of the issues inherent to monitoring data raised in \citet{buckland2017}, such as the scarcity of data, low representability and small geographical scale. 
This opens new possibilities for more robust international assessments of species distributions and abundance using count data from diverse national monitoring programs, which is of paramount importance for understanding global change impacts on biodiversity \citep{buckland2017,massimino2017}. We validate this framework with a case study aiming at estimating total bird abundance in mid-Scandinavia and a simulation study that explores the effects of mispecification on the proposed models.
\\ \\
This paper is organized as follows: In Section 2, we describe the data from the Norwegian and Swedish monitoring programs in detail. Moreover, we explain how we preprocessed these census data, present an exploratory analysis and introduce the set of candidate explanatory variables for our models. In Section 3, models as well as inference methodology and measures for evaluating and comparing models are presented. In Section 4, we set up a simulation study to explore how the proposed models perform in scenarios with different  relation between the observed and the true abundances. In Section 5, results of both the simulation study and the case study using bird counts in mid-Scandinavia are presented. The paper finishes in Section 6 with the discussion of the results and concluding remarks.

%% file: data.tex
\section{Bird monitoring surveys data}
	
\subsection{TOV-E and BBS data}
	
The Norwegian common bird monitoring scheme (TOV-E), coordinated by the Norwegian Institute for Nature Research (NINA) and the Norwegian Ornithological Society (NOF) since 2006, was established to monitor population variation for common breeding terrestrial birds on a national scale in a representative way. Surveys (i.e. count of pairs of birds of all observed species) are carried out by experienced ornithologists that follow a standardized protocol \citep{Kalas2002}. Each census route  (n = 492) contains between 12 and 20 (average = 18.8) point counts 300 m apart describing a square (see Fig. 1) with side = 1.5 km (deviation of this shape are allowed and recorded when the geographic/topographic conditions do not allow the observer to walk, e.g. sea/lakes, glaciers, rough mountainous terrain). A total of 229 species are heard or seen at the entirety of the point counts of TOV-E during 5 minutes. Approximately 121 of the species are less abundant and/or difficult to detect, so observers are asked to record these species during a line-transect between point counts (see Figure 1 - figure with the configuration of a census site with the twenty points). A random selection of 370 census routes (out of a total of 492 routes across Norway) are visited once a year during the period 20th May to 10th June. TOV-E is designed to cover all relevant habitats throughout the altitudinal and latitudinal gradient in Norway and reports ‘pairs of individuals’ as sampling unit.\\ \\
The Swedish breeding bird survey (hereafter BBS) has been coordinated by Lund University since 1996 and consists of 716 fixed sites across Sweden within a 25-km grid (one route per grid cell, see \citet{Lindstrom2013}). These sites are surveyed once a year between mid-May and mid-June (the breeding period for most bird species in Sweden) though not all sites are surveyed every year (mean = 353 sites per year). The 25-km grid makes sure that the habitats of Sweden are monitored in proportion to their abundance in the country as well as the entire altitudinal and latitudinal gradient where birds are present. At each site, the observer walks an 8-km transect describing a $2\times2$ km square and records all bird species heard and/or seen within 8 h. In addition, the observer has eight 5-min point counts where all birds seen or heard must also be recorded. The point counts take place at each of the corners of the square and at the middle point of the transect (see Fig. 1). Of the circa 250 species breeding in Sweden, 244 are reported in BBS, thus ensuring a good coverage of the breeding birds \citep{Lindstrom2013}. The BBS reports ‘individuals’ as sampling unit, which differs from TOV-E’s reporting unit (pairs; see above).\\ \\
Although these monitoring programs are designed to cover a large part of both countries (Fig. 1), for our case study, we only selected census sites that lie within a polygon defined to produce an approximation of a Gaussian Random Field and make inference about a point pattern in Trøndelag Country, central Norway (see red polygon in Fig. 1, \citep{Lindgren2011,10.1093/biomet/asv064}). This polygon covers a total area of 173.634 $km^2$ and contains 113 census sites in Norway and 70 in Sweden.  The main motivation to reduce the study region from the entire country to a smaller area (defined by the polygon) was strictly computational and for an easier compilation of covariate information. In addition, this region, which is basically within Trøndelag County in central Norway, is largely representative of habitat types, topography and biodiversity found elsewhere in Norway. 

		\begin{figure}[H]
		\center
		\includegraphics[width=1\textwidth]{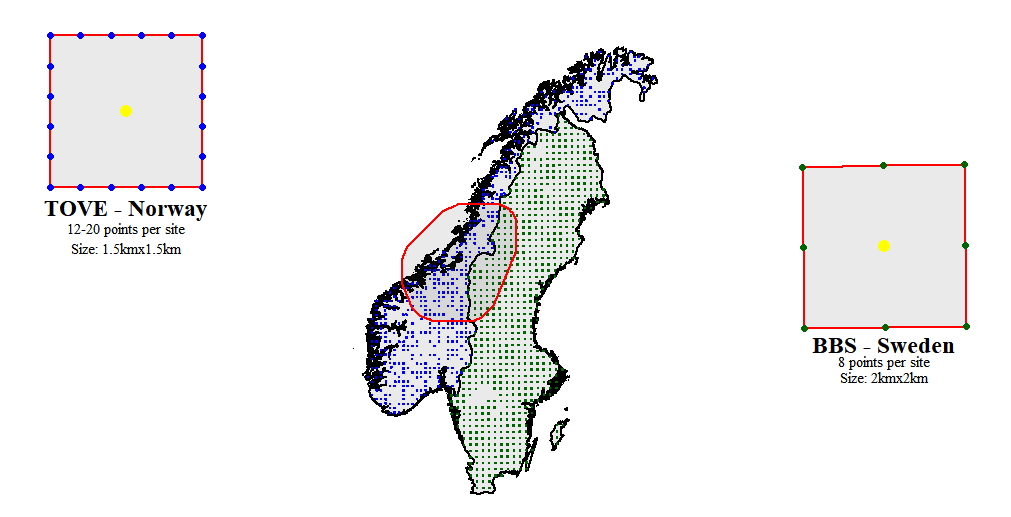} 
		\caption{Spatial location of census sites and sampling points and line transects according to each sampling protocol. Left: Graphical display of sampling protocol of TOV-E census. Blue points: 20 locations for point counts (the number of points vary between 12 and 20 in the different sites). Red lines: Line transects. Yellow point: Centroid associated to each census site (see Section 2.2). Center: Spatial distribution of census sites across Norway (blue sites) and Sweden (green sites). The red polygon represents the study area described in Section 2.1. Right: Graphical display of sampling protocol of BBS census. Green points: 8 locations for point counts. Red lines: Line transects. Yellow point: Centroid associated to each census site (see Section 2.2)  }
		\label{Fig1}
	\end{figure}

\subsection{Exploratory Analysis}

	Our main goal was to develop and validate a new modelling framework to integrate abundance data from standardized monitoring schemes with different sampling protocols. Such a framework can ultimately be used, for example, to detect hotspots of abundance of birds, as in the case we illustrate here (note: we are not interested in the distribution of particular species, but in the distribution of total abundance of birds regardless of the species). In other words, we apply our modeling framework to produce maps of total abundance of birds based on count data from multiple sources - information gathered as part of standardized national bird monitoring schemes in Norway and Sweden that differ in the sampling protocols.\\ \\ 
	The data preparation consisted in averaging across all years (2006 – 2019) the total count of all individuals (regardless of the  species) found at each survey site. That is, we first added up the counts of all individual birds recorded in the points or lines of a given census site and assigned this total count of individuals (regardless of the species present) to the site's centroid (see Figure 1) so that each census site will have one single value of total abundance of birds per year. Next, for each site, we averaged the yearly total abundance of birds across all years that the site was sampled (note: not all sites are censused every year) in the period between 2006 and 2019, so that we ended up with one single value of total abundance of birds per site (temporal average).\\ \\
	Although estimating single-species abundance and distribution maps are commonly used to inform about species of conservation concern, here we wanted to report the total abundance of birds across the region (note: our methodology can also be used to estimate single-species abundances). Estimating total abundance of individuals across a region (as opposed to single-species abundance) has clear implications in spatial conservation planning and prioritization \citep{lehtomaki2013}. For example, \citet{DeLucas2008} estimated the total abundance of raptors in a region to assess the impacts of wind farms on this group of birds. \citet{lindstrom2019} attempted to estimate total density of wading birds across Fennoscandia by combining count data from Norway, Sweden and Finland. However, they did not account for many differences in the sampling protocols. Our modeling framework thus can be applied to account for such differences. Another example of potential use of our method is to get more robust estimates of total abundance of birds to inform authorities and stakeholders where powerlines \citep{Bevanger2014} or wind farms \citep{DeLucas2008} may cause large mortality rates. Although here we present a simplified and more generic analysis (all species have weight = 1, and thus their abundance has the same influence in the resulting map), each species abundance can be multiplied (weighted) by a factor relative to their sensitivity to e.g. powerlines \citep{DAmico2019} so that the resulting map will highlight total abundance hotspots in relation to their sensitivity to the particular issue.\\ \\
	Since we include data from both Norway and Sweden, we explore how the relation of point and line counts differ between surveys from both countries. In Figure 2, we display a scatterplot with the points and line counts at each of the TOV-E (n=113) and BBS (n=70) sites.\\ \\
	\begin{figure}[H]
	\center
	\includegraphics[width=0.8\textwidth]{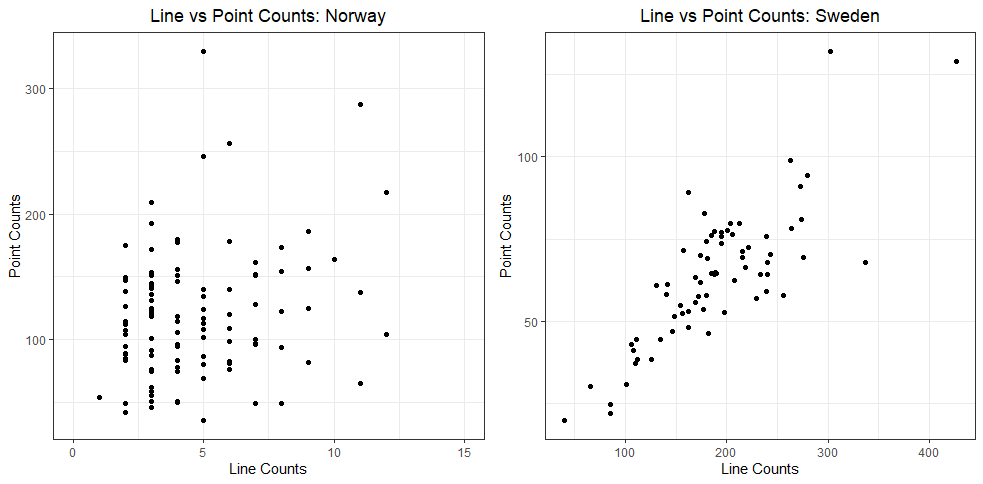} 
	\caption{Scatterplots of line vs point counts in Norway (number of pairs, left) and Sweden (number of individuals, right)}
	\label{explore1}
\end{figure}
	These scatterplots show a linear relation between point and line counts in Sweden, whereas in Norway there is no clear linear association between the counts in points and lines. This is somehow expected due to the census design in Norway, where the line counts are meant to record a reduced subset of species compared to the point counts. This is a common issue, highlighted by \citet{buckland2017} and is often found in many countries when certain species are monitored with special censuses in addition to the general monitoring scheme. Therefore, this is not only an issue when integrating between-countries datasets (e.g. to increase the geographical extent), but also within-country datasets (to increase the representability and number of data points).\\ \\
	
\subsection{Explanatory variables}	

In our case study, we want to apply our new methodology not only to estimate total abundance of birds, but also to produce interpretable estimates of ecological factors associated with it across the region. We have selected three candidate ecological factors that are commonly used in SDMs to explain distribution of birds (e.g. \citet{Bradter2021,Lissovsky2021,Soultan2022}): (i) climatic variables - temperature (average daily temperature from April to July over 2006-2019, downloaded from seNorge.no) and precipitation (average daily precipitation from April to July over 2006-2019, downloaded from seNorge.no), (ii) topography - elevation (Digital Elevation Model at a 10m resolution, DEM10, downloaded from https://kartkatalog.geonorge.no/), and (iii) the land cover surrounding each location expressed as the percentage of each of the following six land covers (urban, mountains, rocky area, water body, forest, and open area) in a square neighbourhood of $2 km \times 2 km$. Land cover information was depicted from the N50 layer (downloaded from https://kartkatalog.geonorge.no/). All rasters files have resolution of $1km \times 1km$ (the elevation data from DEM10 was aggregated to this resolution prior analysis) and are shown in the Supplementary Information.\\ \\
As a first stage of model selection, we computed the correlation coefficient between all the candidate covariates on a fine grid of about 600.000 points. Only one variable in those pairs with $|\rho|>0.7$ was left as a candidate. Those pairs with high correlation were: 1) Elevation and temperature ($\rho=-0.81$). Temperature was discarded; 2) \% of open area and \% of forest ($\rho=-0.83$). \% of open area was discarded.

%% file: models.tex
	\section{Modeling and inference approach}
	
	The specification of our models relies on the assumption that our four sources of observations are obtained from a common underlying ecological process \citep{Milleretal2019}. This assumption arguably makes sense if we consider the fact that national borders of neighbouring countries are not, in general, a key factor for natural changes in biodiversity, although there might be slight differences in conservation policies and governance. Hence, we can assume that a common non-zero mean Gaussian Random Field (GRF) is involved in the generation of the number of individuals at each census site. However, the two different sampling protocols (points and lines), which also differ between the two countries (complementary surveys in Norway and independent surveys in Sweden), result in four groups of counts observed. Moreover, TOV-E counts (Norway) are reported as `number of pairs' of each species whereas BBS counts (Sweden) are reported as `number of individuals' of each species. Therefore, direct inference and comparisons between these four response variables should be made with caution.\\ \\
	The true total bird counts random variable, $Y_{true}(\textbf{s})$ with $\textbf{s} \in D \subset \mathbb{R}^2$ is assumed to follow a Poisson distribution with expected value $\lambda_{true}(\textbf{s})$, expressed as
	
	\begin{equation}
	\log(\lambda_{true}(\textbf{s})) = X^T(\textbf{s})\beta 	 + \omega_1(\textbf{s})
	\label{lambdatrue}
	\end{equation}
	with $X^T(\textbf{s})$ a set of spatial covariates and $\omega_1(\textbf{s})$ a zero-mean GRF that aims at accounting for residual spatial dependency. Both $X^T(\textbf{s})$ and $\omega_1(\textbf{s})$ can include well-established factors that influence variation in the total abundance of birds, in our case study these factors are precipitation and elevation.  We assume a Matérn covariance function for $\omega_1(\textbf{s})$
	\begin{equation}
	\frac{\sigma^2}{\Gamma(\nu)2^{\nu-1}}(\kappa \|s_i-s_j\|)^{\nu} K_{\nu} (\kappa\|s_i-s_j\|)
	\label{MaternCov}
	\end{equation}
	with $\|s_i-s_j\|$ the Euclidean distance between two locations $s_i$, $s_j \in D$. $\sigma^2$ stands for the marginal variance, and $K_{\nu}$ represents the modified Bessel function of the second kind and order $\nu >0$. $\nu$ is the parameter that determines the degree of smoothness of the process, while $\kappa>0$ is a scaling parameter. For $\omega_1(\textbf{s})$, let $\kappa=\kappa_1$,$\nu=\nu_1$ and $\sigma^2 = \sigma_1^2$.\\ \\
	We assume that the observed counts for each sampling protocol are realizations of four random variables conditionally independent given the true abundance, $\lambda_{true}(\textbf{s})$. That is, we assume the four groups of observed counts are realizations of the Poisson random variables:
	
	\begin{align*}
	Y_1(\textbf{s}) &\sim Poisson(\lambda_1(\textbf{s})) \qquad (\text{Point counts in Norway})\\
	Y_2(\textbf{s}) &\sim Poisson(\lambda_2(\textbf{s})) \qquad 
	(\text{Line counts in Norway})\\
	Y_3(\textbf{s}) &\sim Poisson(\lambda_3(\textbf{s})) \qquad
	(\text{Point counts in Sweden})\\
	Y_4(\textbf{s}) &\sim Poisson(\lambda_4(\textbf{s})) \qquad (\text{Line counts in Sweden})	
	\end{align*}
	where $\lambda_j(\textbf{s})$, $j=\{1,2,3,4\}$ are the expected values of the random variables $Y_j(\textbf{s})$. Additionally, we assume $Y_1(\textbf{s}) + Y_2(\textbf{s}) \approx Y_{NO}(\textbf{s})$ as a proxy for total abundance since the line transects are complementary to the point counts in Norway. This assumption does not hold for Sweden since, as mentioned in Section 1, line transects and point counts are regarded as two different independent censuses. In case we wanted to suggest a proxy for the total abundance in Sweden using $Y_3(\textbf{s})$ and $Y_4(\textbf{s})$, we would need to account for a potential overlap (double counting) between the counts observed in points and line transects. Given that we assume a common latent process underlying all the observed abundances, $Y_1(\textbf{s}) + Y_2(\textbf{s})$ works also as a proxy for total abundance of birds in Sweden. This variable is used to produce the predicted total abundance of birds in Section 3. Our final assumption is that there are no differences in observer skills between countries since the census are performed by experienced ornithologists.\\ \\

	\subsection{Models}
	In this section we introduce three model specifications for integrating data from the four sampling protocols introduced in Section 2. Model 1 (see Section 3.1.1) is a model that assumes a linear relation between the expected counts of the four sampling protocols. This is achieved by the introduction of a unique intercept for each sampling scheme. In Section 3.1.2, model 2 is presented. This model allows for a relaxation of the assumption of linear relation between expected counts by incorporating terms that allow to explain any deviation from this assumption through the GRF $\omega_1(\textbf{s})$. Finally, model 3 (see Section 3.1.3) is introduced. This model adds a second GRF, $\omega_2(\textbf{s})$, which aims to account for spatial sources of variation not accounted for in the other parts of the model and not explained by known covariates, \citep{Simmonds2020,selle2020spatial}. It is worth noting that as each of the models proposed depend on $\lambda_{true}(\textbf{s})$, they explicitly account for the factors that influence the variation in abundance.
	\subsubsection{Model 1}
	Based on our exploratory analysis and the four sampling processes present in our dataset, in model 1 we assumed a linear relation between the expected values of the four random variables representing each sampling protocol and $\lambda_{true}(\textbf{s})$. That is, 
	
	\begin{align}
	\lambda_1(\textbf{s}) = \zeta^*_1 \cdot \lambda_{true}(\textbf{s}); \quad \log(\zeta^*_1) &\sim N(0,\tau^*_1)  \nonumber\\
	\lambda_2(\textbf{s}) = \zeta^*_2 \cdot \lambda_{true}(\textbf{s}); \quad \log(\zeta^*_2) &\sim N(0,\tau^*_2) \nonumber\\
	\lambda_3(\textbf{s}) = \zeta^*_3 \cdot \lambda_{true}(\textbf{s}); \quad \log(\zeta^*_3) &\sim N(0,\tau^*_3) \nonumber\\
	\lambda_4(\textbf{s}) = \zeta^*_4 \cdot \lambda_{true}(\textbf{s}); \quad \log(\zeta^*_4) &\sim N(0,\tau^*_4)
	\label{mod1form}
	\end{align}
	with $\zeta^*_j\geq 0$, $j=1,\ldots,4$ the factors that determine the association between the observed and the true counts for each protocol. In real-life problems $\zeta^*_j$ can explain multiple sources of variation that are common to sampling of bird species such as observer differences, observed units, differences in detection probability, among others. The inclusion of this term is also useful to deal with overdispersion \citep{GomezRubio2020}, a common issue when working with count data. In order to avoid identifiability issues, we restate the model in (\ref{mod1form})  in terms of $\lambda_1(\textbf{s})$. That is,
	
	\begin{align}
	\lambda_2(\textbf{s}) = \zeta_2 \cdot \lambda_{1}(\textbf{s}); \quad \log(\zeta_2) &\sim N(0,\tau_2) \nonumber \\
	\lambda_3(\textbf{s}) = \zeta_3 \cdot \lambda_{1}(\textbf{s}); \quad \log(\zeta_3) &\sim N(0,\tau_3) \nonumber \\
	\lambda_4(\textbf{s}) = \zeta_4 \cdot \lambda_{1}(\textbf{s}); \quad \log(\zeta_4) &\sim N(0,\tau_4)
	\label{mod1form1}
	\end{align}
	where $\zeta_j\geq 0$ and $\zeta_j = \frac{\zeta^*_j}{\zeta^*_1}$, $j=\{2,3,4\}$.
	\subsubsection{Model 2}
	In model 2, we relax the assumption of linear relation between the expected value of the number of observed individuals with protocol j, $\lambda_j(\textbf{s})$, and the true intensity, $\lambda_{true}(\textbf{s})$, by including spatial varying terms $(\psi^*_j-1) \cdot \omega_1(\textbf{s})$, $j = \{1,2,3,4\}$. These terms aim to explain any deviation from a linear relation between expected values as a function of a GRF $\omega_1(\textbf{s})$. It is worth noting that model 1 (see above) is a special case of model 2 with $\psi^*_j=1$. We define model 2 as:
	
	\begin{align}
	\lambda_1(\textbf{s}) = \zeta^*_1 \cdot \lambda_{true}(\textbf{s}) \cdot \exp\{(\psi^*_1-1)\cdot \omega_1(\textbf{s})\}; \quad \log(\zeta^*_1) &\sim N(0,\tau^*_1)  \nonumber\\
	\lambda_2(\textbf{s}) = \zeta^*_2 \cdot \lambda_{true}(\textbf{s}) \cdot \exp\{(\psi^*_2-1)\cdot \omega_1(\textbf{s})\}; \quad \log(\zeta^*_2) &\sim N(0,\tau^*_2) \nonumber\\
	\lambda_3(\textbf{s}) = \zeta^*_3 \cdot \lambda_{true}(\textbf{s})\cdot \exp\{(\psi^*_3-1)\cdot \omega_1(\textbf{s})\}; \quad \log(\zeta^*_3) &\sim N(0,\tau^*_3) \nonumber\\
	\lambda_4(\textbf{s}) = \zeta^*_4 \cdot \lambda_{true}(\textbf{s})\cdot \exp\{(\psi^*_4-1)\cdot \omega_1(\textbf{s})\}; \quad \log(\zeta^*_4) &\sim N(0,\tau^*_4)
	\label{mod2form}
	\end{align}
	Again, to avoid identifiability issues, we restate the model in (\ref{mod2form}) in terms of $\lambda_{1}(\textbf{s})$ as: 
	\begin{align}
	\lambda_2(\textbf{s}) = \zeta_2 \cdot \lambda_{1}(\textbf{s})\cdot \exp\{(\psi_2-1) \cdot\omega_1(\textbf{s})\}; \quad \log(\zeta_2) &\sim N(0,\tau_2) \nonumber \\
	\lambda_3(\textbf{s}) = \zeta_3 \cdot \lambda_{1}(\textbf{s}) \cdot \exp\{(\psi_3-1) \cdot \omega_1(\textbf{s})\} ; \quad \log(\zeta_3) &\sim N(0,\tau_3) \nonumber \\
	\lambda_4(\textbf{s}) = \zeta_4 \cdot \lambda_{1}(\textbf{s}) \cdot \exp\{(\psi_4-1) \cdot \omega_1(\textbf{s})\}; \quad \log(\zeta_4) &\sim N(0,\tau_4)
	\label{mod2form1}
	\end{align}
	In the scales of the linear predictors in (\ref{mod2form}) , $\psi_j=\psi^*_j - \psi^*_1 + 1$, $j=\{2,3,4\}$ are scaling coefficients for the common GRF, $\omega_1(\textbf{s})$, in each likelihood. They quantify to what extent the departure of the assumption of linearity is explained by $(\psi^*_j-1) \cdot \omega_1(\textbf{s})$.  In real-life scenarios this departure can be related with sources of variation with spatial structure such as differences in detectability, among others. Therefore, we would expect posterior densities for $\psi_3$ and $\psi_4$ to be around 1 in our case study. While for $\psi_2$ we expect different results because line and point counts in Norway do not seem to follow a linear relation (see Section 2; Fig. 2). Due to the different characteristics of line transect surveys in Norway, we propose model 3.
	
		\subsubsection{Model 3}
		
In addition to causing departure from a linear relation between true and observed counts, species detectability may also change with the census technique used (i.e. one of our data sources, the line transects in TOV-E, targeted only a subset of species as it is regarded as a complementary survey to the point counts). Hence, in model 3 we included a second GRF, $\omega_2(\textbf{s})$ to try to account for the characteristics of this observation process. In case that no explanatory variable that explain the particular characteristics of the sampling protocol is available, a second GRF can be added as a way to account for them, \citep{Simmonds2020}. This is included as an additive term in the linear predictor, as follows:
	
		\begin{align}
	\lambda_2(\textbf{s}) &= \zeta_2 \cdot \lambda_{1}(\textbf{s})\cdot \exp\{(\psi_2-1)\omega_1(\textbf{s})\} \cdot \exp\{\omega_2(\textbf{s})\} \nonumber \\
	\lambda_3(\textbf{s}) &= \zeta_3 \cdot \lambda_{1}(\textbf{s}) \cdot \exp\{(\psi_3-1)\omega_1(\textbf{s})\} \nonumber \\
	\lambda_4(\textbf{s}) &= \zeta_4 \cdot \lambda_{1}(\textbf{s}) \cdot \exp\{(\psi_4-1)\omega_1(\textbf{s})\}
	\end{align}
We assume a Matérn covariance function as in (\ref{MaternCov}) for $\omega_2(\textbf{s})$, with parameters $\kappa=\kappa_2$, $\nu=\nu_2$ and $\sigma^2 = \sigma_2^2$
	\subsubsection{Prior specification}
	
	For the GRFs $\omega_k(\textbf{s})$, $k=\{1,2\}$, the parameters $\nu_k$ in the Matérn covariance function are fixed to be 1. The interest is put on the spatial ranges $\rho_k$ and on the standard deviation of the GRFs, $\sigma_k$. $\rho_k$ are related to $\kappa_k$ through $\rho_k=\sqrt{8}/\kappa_k$. The prior distributions of these two parameters are specified by making use of Penalized Complexity (PC) priors, \citep{doi:10.1080/01621459.2017.1415907}. In this case, we set $P(\rho_1<20000)=0.1$ and $P(\sigma_1>1)=0.1$ for $\omega_1(\textbf{s})$, while $P(\rho_2<2000)=0.1$ and $P(\sigma_2>3)=0.1$ for $\omega_2(\textbf{s})$. This means, for example, that under this prior specification, a standard deviation greater than 1 is regarded as large, while a spatial range below 20 kilometers is considered unlikely for $\omega_1(\textbf{s})$.  The parameters in $\boldsymbol{\beta}$ have Normal prior with mean 0 and precision 0.01. Let $\log(\zeta_j)\sim N(0,\tau_j)$, $j=\{2,3,4\}$, where the logarithm of each $\tau_j$ has a log-Gamma prior with parameters 1 and 0.00005. For the parameters $\psi_j$, $j={2,3,4}$ in models 2 and 3, we set a normal prior with mean 1 and precision 0.1.\\ \\
	We have now defined a group of three candidate models. In the upcoming subsections, we introduce the methodological approach for fitting them and for selecting a model that suits best for our problem.
	\subsection{Inference and computational approach}
	
	The models introduced in Section 3.1 were fitted making use of the Integrated Nested Laplace Approximation (INLA), \citep{Rue2009} and the Stochastic Partial Differential Equation (SPDE) approach \citep{Lindgren2011}. INLA is a faster alternative to Monte Carlo Markov Chains (MCMC) for performing Bayesian inference for latent Gaussian models. INLA aims at producing a numerical approximation of the marginal posterior distribution of the parameters and hyperparameters of the model. Further details can be found in \citet{Rue2009} and \citet{blangiardo2015spatial}. Since we deal with continuous spatial processes in our models, the SPDE approach emerges as an efficient representation of $\omega_1(\textbf{s})$ and $\omega_2(\textbf{s})$. It is based on the solution of a SPDE which can be approximated through a basis function representation defined on a triangulation of the spatial domain. More details are available in \citet{Lindgren2011} and \citet{blangiardo2015spatial}. \\ \\ 
	
	\subsection{Assumptions and possible extensions}
	
	This new modelling framework is developed to integrate count data collected in designed surveys that follow different standardized protocols. Particularly, in the case study presented here, the bird surveys introduced in Section 2 are designed to minimize biases due to variation in the time of sampling or observer expertise. For this reason, the models presented in our case study assume, in principle, that these external sources of variation that could affect the observation process are constant across sites or negligible. However, these models are flexible enough to explicitly account for factors that may affect the observation process of each sampling protocol, and can thus be accounted for.\\ \\
	There may be, however, other potential sources of variation when working with monitoring data, which also depend on the taxon being surveyed. Hence, as mentioned in Section 3, our method includes relevant terms for quantifying the effect of potential sources of noise in the observation process. Our models incorporate the terms $\zeta$ to explain what proportion of the true abundance is explained by each of the observation processes. That is, $\zeta_j$  quantifies the effect of each sampling protocol on the observed abundances. This effect comprises sources of variation such as differences in the observed units, differences in detectability and potential differences in the expertise of the observers. In many real-life scenarios these terms do not provide enough quantification of the effect of the sampling protocols as there are sources of variation in the sampling process that have spatial variation that cannot be summarized in one term. Therefore, the Gaussian Random Field that drives the true abundance (in our case study, the total abundance of birds) or a second GRF is also used to account for sources of variation that have a spatial behaviour.\\ \\
	This modeling framework also allows to explicitly account for factors that affect the observation process of each sampling protocol. To show how this can be done, we take model 2 as our reference to explicitly account for a factor that influences the observed number of individuals. We now assume that unlike our case study, there are several factors affecting the observed total abundance of birds. As seen in equation (\ref{mod2form1}) in Section 3.1.2., the term $\zeta_j \cdot \exp \bigg\{ (\psi_j-1) \omega_1(\textbf{s}) \bigg\}$  accounts for the effect the sampling protocol j has on the observed abundance. In addition to the spatial effect driven by $\omega_1(\textbf{s})$,  the term $\zeta_j$ can be further explained, for example, by a fixed effect $z$ as follows:
\begin{equation}
    \zeta_j = \alpha_{0j} + \alpha_{1j}z
    \label{eqzeta}
\end{equation}
This is a straightforward way to explicitly account for multiple factors that may influence the observation process of the sampling protocol $j$. Factors with a spatial or temporal structure can be accounted for through random effects with these structures.\\ \\
Given the additional parameters to be estimated and the increased complexity of the model when the effect of these factors are accounted for explicitly, structural identifiability issues may arise. Therefore, in order to overcome these issues, it is recommended to constrain the parameters in (\ref{eqzeta}). This can be achieved by either having additional data that inform on these factors or informative prior information of the parameters involved in (\ref{eqzeta}).\\ \\
Acquiring additional data to account for factors that affect the observation process of each sampling protocol might be possible by integrating data, for example, from schemes with sampling protocols designed to gather information on species detection probabilities through repeated visits to the sites or distance sampling \citep{jarvinen1983,Milleretal2019}. In our case study, the temporal variation of birds is not considered to compute the total abundance of birds across the study region. Rather, this temporal variation is removed by averaging the total count of birds at each site over the 14 years (2006 - 2019). This is also a convenient assumption as we do not have information (counts) at every census site every year (i.e. not all sites are surveyed every year). Furthermore, we believe that the overall state of important sites for birds has remained similar in the past 14 years (i.e. bird-rich areas in 2006, at the beginning of the monitoring scheme are still bird-rich areas in 2019, even if the species composition might have changed slightly).\\ \\

	\subsection{Model assessment}
	
	In order to assess and compare competing models such as the ones we are fitting in upcoming sections, we employed the Deviance Information Criterion (DIC), \citep{doi:10.1111/1467-9868.00353}, the Watanabe-Akaike Information Criterion (WAIC), \citep{Watanabe:2010:AEB:1756006.1953045}, the logarithm of the pseudo marginal likelihood (LPML) \citep{blangiardo2015spatial} and the Continuous Rank Probability Score (CRPS) \citep{Gneiting2007}. \\ \\
	DIC makes use of the deviance of the model
	\begin{equation*}
	D(\theta) = -2 \log(p(\textbf{y}|\boldsymbol{\theta}))
	\end{equation*}
	to compute the posterior mean deviance $\bar{D}=E_{\boldsymbol{\theta}|\textbf{y}}(D(\boldsymbol{\theta}))$. In order to penalize the complexity of the model, the effective number of parameters,
	\begin{equation*}
	p_D = E_{\boldsymbol{\theta}|\textbf{y}}(D(\boldsymbol{\theta})) - D(E_{\boldsymbol{\theta}|\textbf{y}}(\boldsymbol{\theta})) = \bar{D} - D(\bar{\boldsymbol{\theta}})
	\end{equation*}
	is added to $\bar{D}$. Thus, 
	\begin{equation*}
	DIC = \bar{D} + p_D.
	\end{equation*}
	The Watanabe-Akaike Information Criterion is based on the posterior predictive density, which makes it preferable to the Akaike and the deviance information criteria, since according to \citet{Gelman2014} it averages over the posterior distribution rather than conditioning on a point estimate. It is empirically computed as
	\begin{equation*}
	-2\bigg[\sum_{i=1}^{n} \log\bigg(\frac{1}{S} \sum_{s=1}^{S} p(y_i|\theta^s) \bigg) + \sum_{i=1}^n V_{s=1}^S ( \log p(y_i|\theta^s)) \bigg]
	\end{equation*}
	with $\theta^s$ a sample of the posterior distribution and $V_{s=1}^S$ the sample variance.\\\\
	Another criterion to compare the models is LMPL,defined as:
	\begin{equation*}
	LPML = \sum_{i=1}^n \log ({CPO}_i)
	\end{equation*}
	It depends on $CPO_i$, the Conditional Predictive Ordinate at location $\textbf{s}_i$, \citep{doi:10.1111/j.2517-6161.1990.tb01780.x}, a measure that assesses the model performance by means of leave-one-out cross validation. It is defined as:
	\begin{equation*}
	CPO_i = p(y_i^*|y_f)
	\end{equation*}
	with $y_i^*$ the prediction of $y$ at location $\textbf{s}_i$ and $y_f = y_{-i}$.\\ \\
	Lastly, we will compare the predictive performance of our models using the Continuous Rank Probability Score (CRPS). It makes possible to compare the estimated posterior mean and our observed values while accounting for the uncertainty of the estimation, \citep{Gneiting2007,selle2019}. It is defined as:
	
	\begin{equation*}
	CRPS(F,y) = \int_{-\infty}^{\infty} (F(u)-1\{y\leq u\})^2
du	\end{equation*}
with $F$, the cumulative distribution of the estimated posterior mean, and $y$ is the observed value. The smaller CRPS is, the closer the estimated value is to the observed one.

%% file: simulation.tex
    	\section{Simulation studies}
	
	We set up three simulation studies based on the case study of total abundance of birds in mid-Scandinavia that allow us to assess the performance of the models proposed in Section 3, when the true data generating model either assumes linear relation between the counts (Scenario 1), deviates from this assumption due to some spatial factor explained by a GRF (Scenario 2) or when one group of observed counts is considerably affected by additional spatial sources of variation (Scenario 3).  We used the same sites as the observations in the TOV-E and BBS surveys (Fig. 1). To start, we simulated the true intensity, $\lambda_{true}(\textbf{s})$ as:
	\begin{equation*}
	\log(\lambda_{true}(\textbf{s})) = \beta_0 + \beta_1 PREC(\textbf{s}) + \omega_1(\textbf{s})
	\end{equation*}
	with $PREC(\textbf{s})$, the precipitation at location $\textbf{s}$ in the study region (see Figure S.1.), and $\omega_1(\textbf{s})$ a GRF with range $\rho = 15 km$ and $\sigma^2 = 0.14$. Further, we specified $\beta_0=4.70$ and $\beta_1=-0.20$. These values were chosen based on the posterior marginal distribution of these parameters in the real-data application. Next, we simulated observations representing the surveys, i.e. using four different Poisson models with parameters $\lambda_j(\textbf{s})$, $j=\{1,\ldots,4\}$.Table \ref{simuscenarios} summarizes the two simulation scenarios proposed for $\lambda_j(\textbf{s})$\\ \\
	\begin{table}[H]
		\centering
	
		\begin{tabular}{cc}
			\multicolumn{1}{l}{Scenario} & Simulated $\lambda_j(\textbf{s})$ \\
			\midrule
			1     & $\lambda_j(s) = \zeta^*_j \cdot \lambda_{true}(\textbf{s})$\\
				 \\
			2     & $\lambda_j(s) = \zeta^*_j \cdot \lambda_{true}(\textbf{s}) \cdot \exp((\psi^*_j - 1) \cdot \omega_1(\textbf{s}))$ \\ \\
    \multirow{2}[0]{*}{3} & $\lambda_j(s) = \zeta^*_j \cdot \lambda_{true}(\textbf{s}) \cdot \exp((\psi^*_j - 1) \cdot \omega_1(\textbf{s})); j=\{1,3,4\}$ \\
          & $\lambda_2(s) = \zeta^*_2 \cdot \lambda_{true}(\textbf{s}) \cdot \exp((\psi^*_2 - 1) \cdot \omega_1(\textbf{s})+\omega_2(\textbf{s}))$  \\
		\end{tabular}%
		\caption{Simulation scenarios}
		\label{simuscenarios}%
	\end{table}%
	For each scenario we simulated 100 datasets with $\zeta^*_1=0.91$, $\zeta^*_2=0.04$, $\zeta^*_3=0.57$ and $\zeta^*_4=1.72$. While we assume a linear relation between $\lambda_j(\textbf{s})$ and $\lambda_{true}(\textbf{s})$ in Scenario 1, in Scenarios 2 and 3 the relation between $\lambda_j(\textbf{s})$ and $\lambda_{true}(\textbf{s})$ is assumed to follow (\ref{mod2form}) with $\psi^*_1 = 1$, $\psi^*_2 = 1.57$, $\psi^*_3 = 1.09$ and $\psi^*_4 = 1.21$. These settings are based on the posterior marginal distribution of the parameters in the real data case study (presented in Section 5.2). The three simulation scenarios closely mimicked real data application by making two of the simulated counts only observed in Norway, and the other two only observed in Sweden. For each simulated dataset we fitted the three models proposed in Section 3. A second group of simulation scenarios was proposed by taking more extreme values of the posterior marginal distributions. The results and more details on this simulation scenario are  
	discussed in Section 5.1 and the supplementary information.
	 \\ \\
To assess the performance of each model in each scenario, we simulated 10000 realizations $\{\theta^p_{jkl}\}, j=,1\ldots,10000$, from the posterior distribution of each parameter $\theta$ for dataset $k =1,\ldots,100$ in scenario $l = 1,2,3$. Thus, the mean bias and the Root Mean Square Error (RMSE) for dataset $k$ in scenario $l$ are computed as:
\begin{align*}
bias_{kl} &= \frac{1}{10000} \sum_{j=1}^{10000} \big(\theta^p_{jkl}-\tilde{\theta} \big)\\ \\
RMSE_{kl} &= \large \sqrt{\frac{1}{10000} \sum_{j=1}^{10000} \big(\theta^p_{jkl}-\tilde{\theta} \big)^2}
\end{align*}
with $\tilde{\theta}$ the true value of the parameter $\theta$.

%% file: results.tex
\section{Results}
\subsection{Simulation Studies}
\begin{figure}[H]
	\center
	\includegraphics[width=0.8\textwidth]{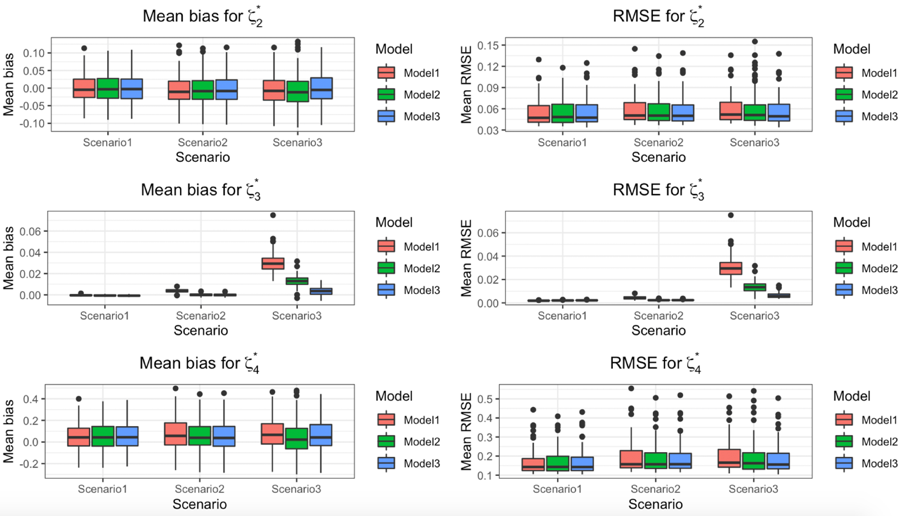} 
	\caption{Mean bias (left) and RMSE (right) for parameters $\zeta^*_2$ (upper panels), $\zeta^*_3$ (central panels) and $\zeta^*_4$ (lower panels) ) for each model in simulation scenario 1 (assumption of linear relationship between expected abundances), scenario 2 (non-linear relation between expected abundances explained by $\omega_1(\textbf{s})$) and scenario 3 (an extra spatial source of variation affecting only one of the groups of observed counts). }
	\label{simures1}
\end{figure}

The 100 datasets generated in each of the proposed scenarios were fitted using the three proposed models in Section 3 and the results summarized here using the measures of performance introduced in Section 4. We only show the mean bias and RMSE for the parameters $\zeta^*_2$, $\zeta^*_3$ and $\zeta^*_4$ as they are key to understand how the different response variables interact with each other (Fig. \ref{simures1}).\\ \\ 
Figure \ref{simures1} shows that the estimation of the proportional relation between the four likelihoods performed similarly for the three models when the truth is that the four likelihoods are linearly related (Scenario 1). Model 1 (which assumes linear relationship between the expected counts) performed, as expected, slightly better than the other two models as this is the model that generated the datasets. However, when we introduced some deviation from the assumption of linearity in our data generating process (Scenario 2), model 1 underperformed relative to the other two models. This is true for the three parameters of interest (Figure \ref{simures1}). Models 2 and 3 performed better in terms of bias and RMSE, whereas the estimates produced by model 1 were biased and showed higher variability. Lastly, when an additional source of variation affected only one of the likelihoods (Scenario 3), the three models performed similarly as in Scenario 2, except for the hyperparameter $\zeta^*_3$, which is part of the likelihood affected by the extra source of variation. For this hyperparameter, the differences in performance between the three models increased considerably as model 3 produced less biased and variable estimates of this hyperparameter.\\ \\

\begin{table}[H]
  \centering
    \begin{tabular}{c c c c c c c c c c}
    \toprule
    Scenario & Model & \multicolumn{2}{c}{$\beta_0$} & \multicolumn{2}{c}{$\beta_1$} & \multicolumn{2}{c}{$\rho$(km)} & \multicolumn{2}{c}{$\sigma$} \\
    \midrule
    \multicolumn{1}{c}{} &       & Bias  & RMSE  & Bias  & RMSE  & Bias  & RMSE  & Bias  & RMSE \\
    \midrule
    \multirow{6}[12]{*}{1} & \multirow{2}[4]{*}{1} & -0.112 & 0.120 & $1.62 \cdot 10^{-3}$ & 0.040 & -1.567 & 4.729 & 0.076 & 0.097 \\
\cmidrule{3-10}          &       & (0.037) & (0.035) & (0.028) & (0.011) & (3.982) & (1.946) & (0.080) & (0.076) \\
\cmidrule{2-10}          & \multirow{2}[4]{*}{2} & -0.116 & 0.125 & $2.74\cdot 10^{-3}$ & 0.043 & -1.380 & 4.812 & 0.115 & 0.129 \\
\cmidrule{3-10}          &       & (0.036) & (0.034) & (0.028) & (0.011) & (4.330) & (2.267) & (0.096) & (0.099) \\
\cmidrule{2-10}          & \multirow{2}[4]{*}{3} & -0.120 & 0.129 & $1.48\cdot10^{-4}$ & 0.044 & -1.497 & 4.762 & 0.122 & 0.135 \\
\cmidrule{3-10}          &       & (0.038) & (0.035) & (0.030) & (0.011) & (3.978) & (2.021) & (0.088) & (0.093) \\
    \midrule
    \multirow{6}[12]{*}{2} & \multirow{2}[4]{*}{1} & -0.112 & 0.119 & $1.37\cdot10^{-3}$ & 0.040 & -1.132 & 4.681 & 0.066 & 0.089 \\
\cmidrule{3-10}          &       & (0.037) & (0.035) & (0.028) & (0.011) & (4.152) & (2.057) & (0.070) & (0.065) \\
\cmidrule{2-10}          & \multirow{2}[4]{*}{2} & -0.111 & 0.120 & $1.05\cdot10^{-4}$ & 0.038 & -0.880 & 4.704 & 0.069 & 0.091 \\
\cmidrule{3-10}          &       & (0.036) & (0.034) & (0.024) & (0.009) & (4.160) & (2.148) & (0.070) & (0.065) \\
\cmidrule{2-10}          & \multirow{2}[4]{*}{3} & -0.104 & 0.113 & $-1.92\cdot10^{-3}$ & 0.048 & -0.952 & 4.629 & 0.058 & 0.082 \\
\cmidrule{3-10}          &       & (0.038) & (0.035) & (0.049) & (0.025) & (4.013) & (1.981) & (0.067) & (0.060) \\
    \midrule
    \multirow{6}[12]{*}{3} & \multirow{2}[4]{*}{1} & -0.112 & 0.120 & $1.42\cdot10^{-3}$ & 0.040 & -1.056 & 4.652 & 0.063 & 0.087 \\
\cmidrule{3-10}          &       & (0.037) & (0.035) & (0.028) & (0.011) & (3.927) & (1.972) & (0.069) & (0.065) \\
\cmidrule{2-10}          & \multirow{2}[4]{*}{2} & -0.111 & 0.120 & $5.06\cdot10^{-5}$ & 0.038 & -0.596 & 4.783 & 0.069 & 0.089 \\
\cmidrule{3-10}          &       & (0.036) & (0.034) & (0.024) & (0.009) & (4.288) & (2.139) & (0.070) & (0.065) \\
\cmidrule{2-10}          & \multirow{2}[4]{*}{3} & -0.112 & 0.119 & $-2.05\cdot10^{-4}$ & 0.040 & -1.837 & 5.024 & 0.088 & 0.107 \\
\cmidrule{3-10}          &       & (0.037) & (0.035) & (0.028) & (0.011) & (4.240) & (2.064) & (0.079) & (0.076) \\
    \bottomrule
    \end{tabular}%
\caption{Mean bias and RMSE for parameters $\beta_0$, $\beta_1$, $\rho$ and $\sigma$ in simulation scenario 1 (assumption of linear relationship between expected abundances), scenario 2 (non-linear relation between expected abundances explained by  $\omega_1(\textbf{s})$) and scenario 3 (an extra spatial source of variation affecting only one of the groups of observed counts). In parentheses, the standard error of each performance measurement.}
  \label{simubiasrmse}%
\end{table}%

\begin{figure}[H]
	\center
	\includegraphics[width=0.8\textwidth]{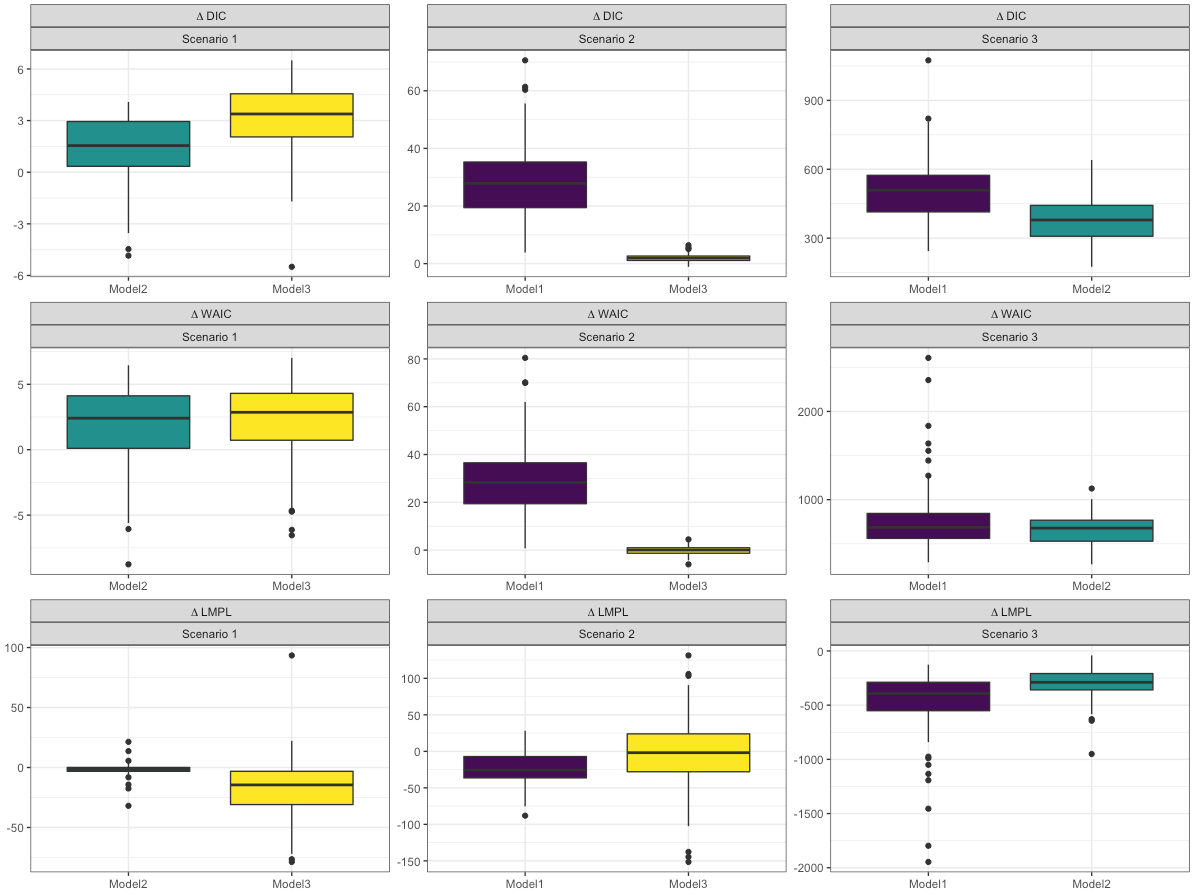} 
	\caption{Differences in DIC, WAIC and LMPL between the model that generated the observed counts in each simulation scenario (Scenario 1, generated according to model 1;scenario 2, generated according to model 2 and scenario 3, generated according to model 3)  and the other two models proposed in Section 3. }
	\label{simurescritcomp}
\end{figure}

Our results show that there are only marginal differences in the fixed effects $\beta_0$ and $\beta_1$ between the three models in all the scenarios. However, larger differences are observed for the hyperparameters of $\omega_1(\textbf{s})$. For example, in the three scenarios the bias of $\rho$ was smaller for model 2 compared to the other two models, but at the same time it produced estimates of $\rho$ with larger RMSE than the other two models.\\ \\
In this simulation study we have also explored the selection of the best model according to the comparison criteria DIC, WAIC and LMPL (See Section 3.4). For each scenario, we computed the differences in each criterion between the model that generated the 100 datasets of the scenario and the other two models. The summaries of these differences are displayed in Figure \ref{simurescritcomp}.

Figure \ref{simurescritcomp} shows small differences in DIC and WAIC between the three models when model 1 generates the observed counts (Scenario 1). In Scenario 1, the predictive performance, measured by LMPL, was similar for model 1 (the one that generated that data) and model 2, while model 3 underperformed. In Scenario 2, model 2 (generating model) and model 3 performed similarly based on all performance comparisons, but model 1 underperformed considerably. In scenario 3, where the observed counts are generated according to a more complex specification (i.e. one sampling protocol is affected by an additional source of variation), model 3 had better goodness of fit and predictive performance with large differences in DIC, WAIC and LMPL with respect to the other two models. The difference in performance between models increases as the complexity of the data generating process increases (Fig. \ref{simurescritcomp}). \\ \\
The results for the second group of simulations can be found in the Supplementary Information. Results match those obtained with the first group of simulations above. In Scenario 1, all three models perform similarly. As in the first simulation study, when the complexity of the model that generates the data increases, models 2 and 3 outperform model 1. Nevertheless, unlike in the first simulation study, model 2 outperformed model 3 in Scenario 3 (generated by model 3) for $\zeta_3$ as it produced less biased estimates.

\subsection{Results of the case study on total abundance of birds in mid-Scandinavia}

We fitted our three models (see Section 3) to count data from the common bird monitoring schemes in Norway and Sweden (see Section 2) to estimate total abundance of birds across mid-Scandinavia with precipitation and elevation as explanatory variables. These two were selected from all the variables considered \textit{a priori}, as it was the subset of candidate variables that produced the best results in terms of goodness of fit (see Supplementary Information for an overview of the performance of other competing models).  The most demanding model in terms of computation time was model 3, which run in 60 seconds. In Table 3, we report the posterior mean, standard deviation and quartiles of the most relevant parameters from the three models.


\begin{sidewaystable}
	\scriptsize			
	\centering
	\begin{tabular}{clllllllllllllll}
		\cmidrule{2-16}    \multicolumn{1}{r}{} & \multicolumn{15}{c}{Model} \\
		\cmidrule{2-16}    \multicolumn{1}{r}{} & \multicolumn{5}{c}{Model 1}          & \multicolumn{5}{c}{Model 2}          & \multicolumn{5}{c}{Model 3} \\
		\midrule
		Parameter & Mean  & SD    & 0.025q & 0.50q & 0.975q & Mean  & SD    & 0.025q & 0.50q & 0.975q & Mean  & SD    & 0.025q & 0.50q & 0.975q \\
		\midrule
		Intercept & 4.69  & 0.04  & 4.61  & 4.69  & 4.77  & 4.68  & 0.03  & 4.62  & 4.68  & 4.75  & 4.69  & 0.04  & 4.61  & 4.69  & 4.77 \\
		PREC  & -0.12 & 0.04  & -0.19 & -0.12 & -0.04 & -0.20 & 0.03  & -0.26 & -0.20 & -0.14 & -0.11 & 0.04  & -0.18 & -0.11 & -0.04 \\
		ELEV  & -0.29 & 0.04  & -0.38 & -0.29 & -0.21 & -0.39 & 0.04  & -0.46 & -0.39 & -0.32 & -0.27 & 0.04  & -0.35 & -0.27 & -0.19 \\
		$\zeta_2$ & 0.05  & 0.00  & 0.04  & 0.05  & 0.05  & 0.04  & 0.00  & 0.04  & 0.04  & 0.05
		& 0.04  & 0.00  & 0.04  & 0.04  & 0.05 \\
		$\zeta_3$ & 0.51  & 0.03  & 0.45  & 0.51  & 0.57  & 0.48  & 0.03  & 0.43  & 0.48  & 0.54
		& 0.51  & 0.03  & 0.45  & 0.51  & 0.57 \\
		$\zeta_4$ & 1.50  & 0.09  & 1.33  & 1.50  & 1.68  & 1.42  & 0.08  & 1.27  & 1.42  & 1.58
		& 1.50  & 0.09  & 1.32  & 1.49  & 1.69 \\
		$\psi_2$ &  &  &  &  &  & 1.86  & 0.14  & 1.59  & 1.86  & 2.13  
		& 0.61  & 0.16  & 0.30  & 0.61  & 0.91 \\
		$\psi_3$ &  &  &  &  &  & 1.26  & 0.13  & 1.00  & 1.26  & 1.52
		& 1.09  & 0.12  & 0.86  & 1.09  & 1.34 \\
		$\psi_4$ & &  &  &  &  & 1.30  & 0.12  & 1.07  & 1.30  & 1.54  
		& 1.18  & 0.12  & 0.96  & 1.18  & 1.42 \\
		$\rho$ & $1.80 \cdot 10^4$ & $4.00 \cdot 10^3$ & $1.11 \cdot 10^4$ & $1.77 \cdot 10^4$ & $2.68 \cdot 10^4$ &
		
		 $1.80 \cdot 10^4$& $3.88 \cdot 10^3$& $1.17 \cdot 10^4$& $1.75 \cdot 10^4$& $2.69 \cdot 10^4$&
		 
		 $2.01 \cdot 10^4$& $4.12 \cdot 10^3$& $1.29 \cdot 10^4$& $1.98 \cdot 10^4$& $2.90 \cdot 10^4$\\
		$\sigma$& 0.36  & 0.02  & 0.32  & 0.36  & 0.41  & 0.31  & 0.02  & 0.27  & 0.31  & 0.36  & 0.34  & 0.03  & 0.29  & 0.34  & 0.39 \\
		\bottomrule
	\end{tabular}%
	\caption{Posterior mean, standard deviation and quartiles of the most relevant parameters of the models proposed in section 3}
	\label{tovemodtable}%
\end{sidewaystable}%

Table \ref{tovemodtable} shows the associations between precipitation (PREC) and elevation (ELEV) with the expected counts are negative for all the models. The posterior means of the parameters of these two variables have small differences, model 2 estimated stronger association of the explanatory variables (precipitation and elevation) and the response variable (total abundance of birds). The posterior summaries of PREC and ELEV suggest that those locations with higher levels of precipitation and high elevation are expected to have lower total bird counts. The variability and range of the Gaussian field have right skewed posterior distributions based on their posterior medians and means.\\ \\
 \begin{figure}[H]	
 	\center
 	\includegraphics[width=0.8\textwidth]{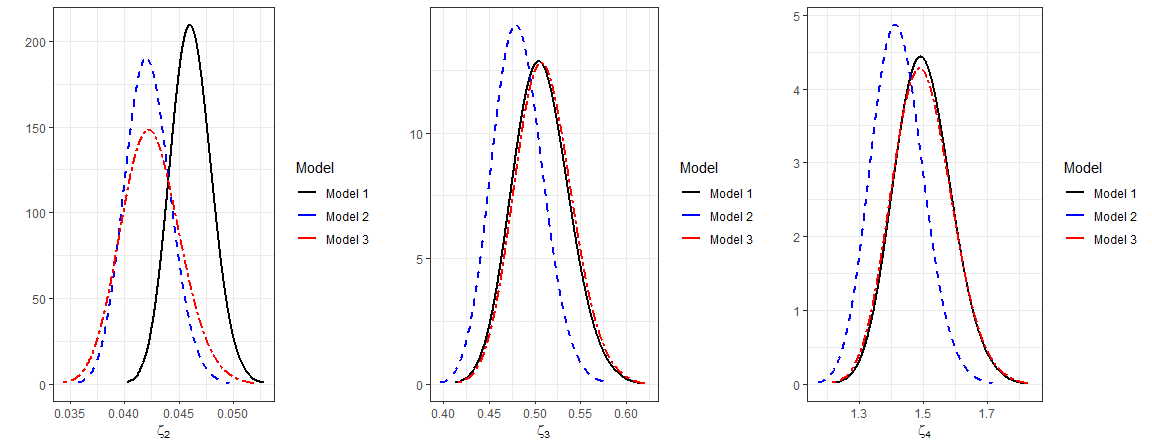} 
 	\caption{Posterior densities of $\zeta_2$ (left), $\zeta_3$ (center) and $\zeta_4$ (right) for each model}
 	\label{zetaspostdist}
 \end{figure}
Figure \ref{zetaspostdist} and Table \ref{tovemodtable} show that the posterior densities of $\zeta_2$ are different between models, with higher posterior mean for model 1 compared to the other models. This result agrees with the exploratory analysis of Section 2, which suggested the necessity of specifying a relaxed linear relationship  between the line and point counts in Norway (linearity was met in Sweden, but not in Norway, see Fig. 2). However, the posterior densities of $\zeta_3$ and $\zeta_4$ are almost identical for models 1 and 3 whereas model 2 estimated posterior distributions for $\zeta_3$ and $\zeta_4$ that are shifted towards lower values (Fig. 5).\\ \\
Large differences in the posterior mean of $\psi_2$ in models 2 and 3 are observed when $\omega_2(\textbf{s})$ is introduced to account for the particularities of the sampling protocol of the line counts in Norway (i.e. in general terms, to account for added complexity due to one of the data collecting protocols considered). While model 2 gives high prevalence to $\omega_1(\textbf{s})$ (posterior mean of $\psi_2=1.90$) as determinant of the departure from linear association, model 3 reduces this prevalence (posterior mean of $\psi_2=0.63$). It arguably means that $\omega_2(\textbf{s})$ accounts for what is particular of this sampling protocol (the added complexity), and what at the same time reduces the leverage of what is shared between this sampling protocol (the line transect in Norway in this case study) and the other protocols. We expect these differences in contribution of $\omega_1(\textbf{s})$ across models to impact their predictive performance. In Figure S.2 we show the posterior mean of $Y_1(\textbf{s})+Y_2(\textbf{s})$, understood as a proxy for the total abundance of birds in our study region (see Section 3). Given the high similarity across mid-Scandinavia, hereafter, we explore the differences in the predicted mean of $Y_1(\textbf{s})+Y_2(\textbf{s})$ between the three models in a smaller sub-region (highlighted with a red square in Figure \ref{postmeanpred}), which encompasses the locations surrounding Trondheimsfjorden and the Norwegian Sea.
 \begin{figure}[H]	
	\center
	\includegraphics[width=1\textwidth]{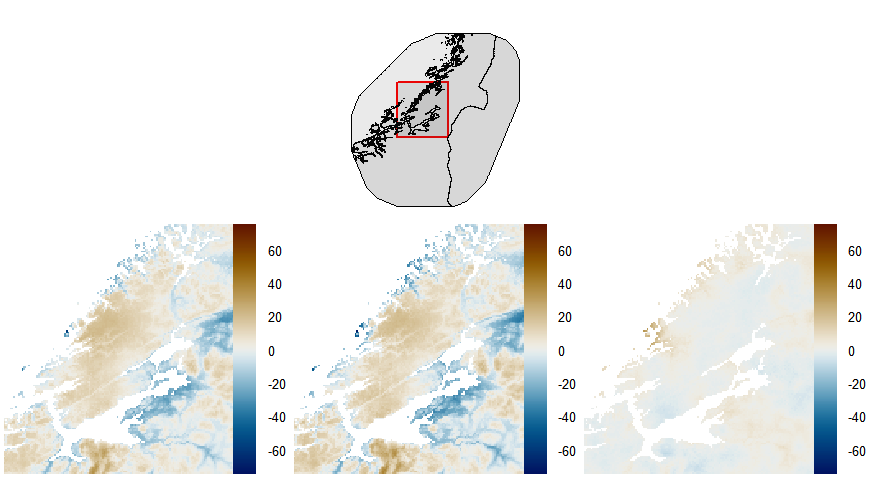} 
	\caption{Top(small): Study region with the red square that encloses the zone chosen for analyzing differences between models. Bottom: Differences in the posterior mean of $Y_1(\textbf{s})+Y_2(\textbf{s})$ (i.e. total abundance of birds) between: model 1 - model 2 (left), model 3 - model 2 (center) and model 1 - model 3 (right)}
	\label{postmeanpred}
\end{figure}
Our three models predicted high total bird counts along the eastern coast of Trondheimsfjorden and on the islands of Hitra and Frøya (Fig. S.9) and low counts at higher elevations such as in the mountainous in the southwest and the north of the study region (Fig S.9.). Model 2 estimates higher counts compared to the other two models along the fjord's coast (dark blue) and lower abundance inland (mainly in the mountains; light brown; Fig. \ref{postmeanpred}). The differences in predicted counts between model 1 and model 3 are smaller (Fig. \ref{postmeanpred}, right panel) compared to those with model 2. However, larger predicted counts are produced by model 3 around the island of Linesøya.\\ \\ 
Our modeling framework allows for computing the uncertainty of our predictions. Here, we assess this by computing the standard error of $Y_1(\textbf{s})+Y_2(\textbf{s})$ (see Fig. \ref{postsdpred} for the standard error of the sub-region highlighted in Fig. \ref{postmeanpred}, and see Fig. S.10 for the standard errors across the entire study region).\\ \\
\begin{figure}[H]	
	\center
	\includegraphics[width=1\textwidth]{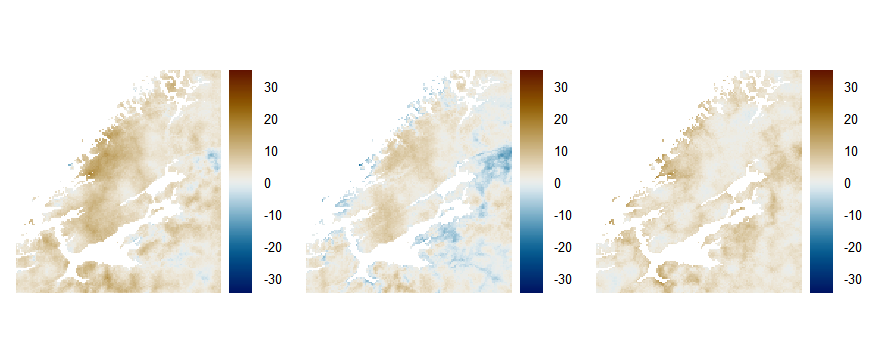} 
	\caption{Differences in the posterior standard error  of $Y_1(\textbf{s})+Y_2(\textbf{s})$ for: model 1 and model 2 (left), model 3 and model 2 (center) and model 1 and model 3 (right)}
	\label{postsdpred}
\end{figure}
The standard error of model 1 is larger than the other two models in most regions (see brown colors, left and right panels in Fig. \ref{postsdpred}). In the zones with higher predicted counts (the coast on the Norwegian Sea and Trondheimsfjorden), model 2 produced predictions with higher uncertainty (dark blue in the central panel), while on the mountains the uncertainty produced by model 3 was larger (light brown in the central panel)
As a way to better appreciate the numerical differences between models, we explored the total predicted counts at the 113 sampling sites in Norway by comparing them against the observed counts (Fig. \ref{predcomp}).
\begin{figure}[H]	
	\center
	\includegraphics[width=1\textwidth]{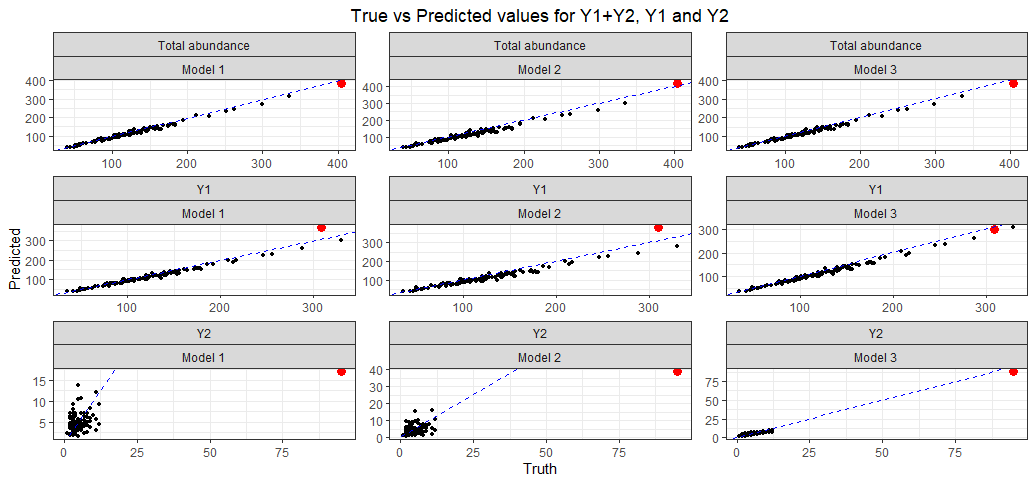} 
	\caption{Comparison of observed vs predicted counts for: total abundance($Y_1(\textbf{s})+Y_2(\textbf{s})$; top row), counts produced via point counts $Y_1(\textbf{s})$ (middle row) and counts produced via line transect counts $Y_2(\textbf{s})$ (bottom row). The performance of model 1, model 2 and model 3 are displayed in the first, second and third column, respectively. A particular site with high total abundance of birds due to presence of gregarious species (in this case) that is only captured by one (the line transect in Norway) of the four census protocols is highlighted in red to allow for a quick assessment of discrepancies between the three models.
}
	\label{predcomp}
\end{figure}

Figure \ref{predcomp} shows the comparison between the predicted and the observed values of total abundance of birds ($Y_1(\textbf{s})+Y_2(\textbf{s})$). Model 1 and model 3 predict very similar values and thus, we also compared the observed and predicted values of the counts gathered via point counts
$Y_1(\textbf{s})$ and line transects $Y_2(\textbf{s})$ separately. Although model 1 and model 3 produce very similar predictions of total abundance of birds, model 3  predicted $Y_1$ and $Y_2$ separately more accurately. This is due to the inclusion of the  GRF  $\omega_2(\textbf{s})$, as it makes it possible to better distribute the abundance between likelihoods and is flexible enough to capture more complex relationships between the census processes. We have highlighted the predicted and observed counts of the site located in the island of Linesøya (in red in Fig. 8) as this is a site where big discrepancies are observed between all the models. Model 1 and model 2 are not able to accurately predict the counts reported in this site by the line transect survey in Norway. This site is a special location where gregarious geese belonging to several species aggregate and form large gaggles (similar examples elsewhere might be sites with (multi-species) colonies, roosting sites or wetlands hosting thousands of waterbirds). Such information is only available if data from several census protocols are combined and properly analysed - our new modelling framework can account for these differences, as our model 3 does in comparison to model 1 that assumes a linear relation.\\ \\
\begin{figure}[H]	
	\center
	\includegraphics[width=0.5\textwidth]{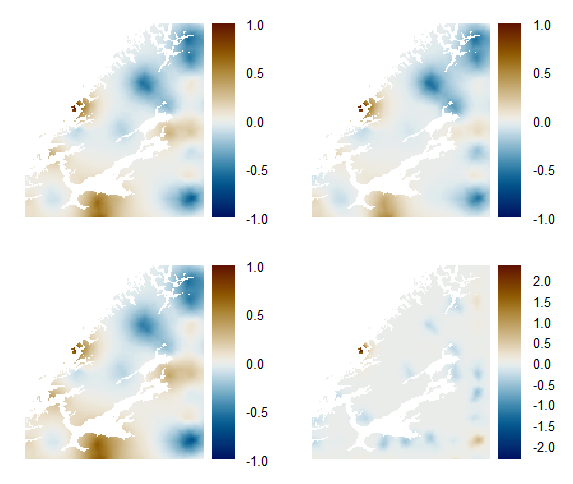}
	\caption{Posterior mean of $\omega_1(\textbf{s})$ for model 1 (upper left), model 2 (upper right) and model 3 (bottom left). Posterior mean of $\omega_2(\textbf{s})$ for model 3 (bottom right)  }
	\label{postgrfs}
\end{figure}
Figure \ref{postgrfs} shows the posterior mean of $\omega_1(\textbf{s})$ for the three models, as well as $\omega_2(\textbf{s})$ for model 3. $\omega_1(\textbf{s})$ is, in general, similar for the three models. The largest difference occurs in $\omega_1(\textbf{s})$ for model 2, which has a shorter spatial range in comparison to the other two models. In addition, the highest contribution of $\omega_2(\textbf{s})$ occurs in Linesøya, an island where high total abundance of birds can be recorded during the line transects, due to high concentrations of geese from several species (see above). Such species form large groups of individuals (so called, gaggles) in some of the islands along the Norwegian coast.\\ \\
Lastly, we compared our three models in terms of goodness of fit and predictive performance (Table 4) using the measures of performance introduced in Section 3.4 and out-of-sample predictive performance measures such as RMSE after brute-force Leave-One-Out Cross Validation (CV), \citep{Vehtari2016} and Leave-One-Site-Out CV. In the former CV scheme, we removed one data point at a time, while for the other we removed both the point and line transect counts. This procedures were computationally demanding, but feasible for our problem as it took 1.76 hours for model 1, 4.2 hours for model 2 and 4.1 hours for model 3. \\ \\
\begin{table}[H]
	\centering
	\begin{tabular}{p{15.835em}p{6.415em}p{6.835em}p{7.585em}}
		\toprule
		Measure of performance & Model 1 & Model 2 & Model 3 \\
		\midrule
		DIC   & 2728.79 & 2751.04 & \textbf{2603.82} \\

		WAIC  & 2876.19 & 2876.96 &	\textbf{2593.69} \\

		RMSE & 165.60 & 145.00 & \textbf{38.63} \\

		LMPL & -1673.50 & -1656.35 & \textbf{-1425.04} \\

		Mean CRPS & 27.21 & 25.93 & \textbf{20.95} \\
		
		RMSE (Leave-One-Site-Out CV) & \textbf{45.70} & 46.39 & 45.88\\
		\bottomrule
	\end{tabular}%
\caption{Measures of performance (see Section 3.3) for models 1, 2 and 3. In bold the model with the best performance.}
	\label{perfmeas}%
\end{table}%
The results show a considerable improvement of the goodness of fit when a second GRF to account for the particular characteristics in one of the observed data sources (line transects in Norway) is added. Moreover, the improvement in predictive performance of model 3 is exemplified by its low values of RMSE for the point count surveys in Norway, its high value of LMPL and its low CRPS for the point transect counts in Norway. The result of the leave-on-site-out CV shows small differences, but model 1 outperformed the other two models.

%% file: discussion.tex
\section{Discussion and conclusions}

The main goal of this paper was to introduce a modeling framework that allows us to model jointly multiple sources of information (count data) that are collected under different sampling protocols. We also presented a simple case study where we used this new methodology to estimate the total abundance of birds in mid-Scandinavia using bird counts in Norway and Sweden. These two countries have well-established bird monitoring programs, but differ in the sampling protocols. Therefore, we proposed a set of models that assumed the same coefficients for the fixed effects in each likelihood and a common GRF. The only difference between the different likelihoods are random intercepts in the linear predictor that aim at accounting for differences in the sampling protocols. For example, while the observed point counts in Norway have pairs of birds as the unit reported, Sweden reports individuals. Having different random intercepts makes possible to establish a proportional relation between the observed counts in the data sources. This is arguably a sensible choice since the biological processes that determine the abundance of species do not generally depend on national borders.\\ \\
Although the assumption of linear relation is reasonable for this case, it is also true that when working with real data allowing for some flexibility with respect to this assumption may correspond better to reality in most cases. This is why we proposed a model that has a common GRF, but with a coefficient that explains how far we are from a linear relation. As seen in the exploratory analysis (Section 2), one of our data sources did not seem to follow the assumption of linear association with the other likelihoods. Hence, we suggested the inclusion of a second GRF to account for the differences of this likelihood. The inclusion  of the second GRF, $\omega_2(\textbf{s})$, was especially useful in our case as we do not have variables at the spatial point level that explicitly inform on the differences of the line count surveys in Norway with respect to the other likelihoods. \citet{Simmonds2020} show the benefits of including an extra GRF to account for sources of bias in the sampling process of Citizen Science data. \\ \\
We assessed the performance of the three models when the key assumptions in the specification of each of them were not met in two simulation studies. The results of these simulations showed that a flexible specification performed similarly to the model that assumed a linear relation (model 1) when the latter model was used to generate the data. On the other hand, when the linear assumption was not met by the data generating model, the gap in performance between models became more evident. This suggests that using the models with flexible specification is always advised, regardless of the nature of the data. The estimates of the parameters in model 1 (the model assuming a linear relation between the observed counts) were biased and more uncertain than the estimates of the same parameters in the other two models. When a more complex scenario was proposed, model 3 (the model with two GRFs) clearly outperformed model 1 and model 2 in every comparison criteria. From the two simulation studies, we can conclude that model 3 is more robust than the other two models to misspecification of the functional form of the model. The parameters that showed higher differences in terms of bias and mean RMSE in the simulation study were the hyperparameters $\zeta_j$. This might be caused by caused by the fact that these parameters are the only ones that are not constrained to be the same for all the likelihoods and therefore they are more sensitive to misspecification. A biased estimate of these hyperparameters might have an impact on the predictions of our models (total abundance of birds, in our case study) as these coefficients can be used as weighting of the different likelihoods when computing the total abundance.	\\ \\
The data of the simulation studies were also used to show why integrating the four sources of information is better for predicting the total counts of birds in more than one country (See Section S.1.2. of the Supplementary Information). We compared the predictive performance of a set of models that include (i) only one of the four sources of information, (ii) two sources of information (from the same country to predict abundance in a given location within the corresponding country - e.g. points and lines from Norway to predict within Norway), and (iii) the four sources of information (points and lines from both countries)(see Table S.2.). The results show that if the goal of the study is to produce predictions in more than one country, then integrating sources of information from both countries is recommended. If the goal of the study is to only produce within-country predictions, then integrating information for more than one country would not provide any additional benefit as the models with two sources of information performs as well as the models with the four sampling protocols.\\ \\
When we applied this methodology to the case study of estimating total bird abundance in mid-Scandinavia, we found some very high counts on the island of Linesøya (compared to elsewhere in the region). This count was recorded during a line transect sampling, which model 1 and model 2 failed to explicitly account for. This is arguably why the differences in goodness of fit between model 1 and model 2 were negligible. The inclusion of a second GRF in model 3 to explain extra complexity (in this case, the line counts in Norway that may produce large number of birds) made sense for our research problem since it was able to explain the large counts in Linesøya, when a large number of geese congregate around these islands. Adding GRFs to the likelihoods in order to account for particularities of each observed response seemed useful and practical in other cases when researchers need to account for complexity that can not be explained with available covariate information. However, this addition should have a clear justification and be applied with caution since giving an ecological interpretation to this random effect may not be a trivial task.\\ \\
Our  modeling framework offers, thus, advantages to integrate data from surveys with different sampling protocols and disjoint spatial locations. In its most simple parametrization it does not explicitly account for any factor that affects the observed total abundance (i.e. detection). For example, in our case study, we have assumed these factors are negligible. However, this modeling framework is flexible enough to explicitly account for factors that influence the observed abundance. As shown in Section 3, these factors can be accounted for by explaining each of the terms $\zeta_j$ in the models proposed as a function of fixed and random effects that affect the observation process. Given the complexity of the models, identifiability issues may arise if the parameters that explain the effect of the factors related to the observation process are not constrained. This issue can be overcome by integrating data that inform on these parameters, or informative prior knowledge about them.\\ \\
The proposed framework  does not explicitly accommodate species-specific characteristics. In our case study it was not necessary as we assumed all the species have the same weight on the estimated total abundance. However, this modeling framework can work for a broader range of goals. 
For example, if one or a group of species are of interest when studying anthropogenic impacts on birds (e.g. total raptor counts \citep{DeLucas2008}), the raw data can be preprocessed according to the purpose of the study. If the goal is to model one species of concern, then getting the subset of the raw data that belong to this species would suffice to apply our methodology and obtain satisfactory results. If, in another case, the question we want to solve is linked to the risk of collision of birds with powerlines (e.g. \citet{DAmico2019}) or rotor blades in wind farms (see \citet{DeLucas2008}), 
we can account for the differences in sensitivity between species (for example soaring raptors, which are proportionally scarce in common bird monitoring schemes, are more sensitive than other bird species). Thus, one would multiply (apply weights) the count of each species in the dataset by a 'species-specific sensitivity factor' to that particular human impact (in this case, counts of raptor species would have a larger weight than other species). Then one would proceed by summing up the new weighted counts to obtain a 'total weighted abundance of birds' at each census site. Our methodology, thus, can provide estimates of such a total weighted abundance across the entire region of interest and maps of 'sensitivity-adjusted hotspots'.
An open question would be then, how to decide the values of these weights, which might be decided based on, for example, expert opinion, traits databases \citep{Tobias2022} and published literature \citep{DAmico2019}. A limitation of this modeling framework is that it lies in the category of purely spatial SDMs and thus it is not possible to explicitly account for any potential temporal variation at small (e.g. within a day) or large (e.g. across years) scale. In our case study this was not a major concern as the temporal span of our data (14 years) is not considered a period in which the distribution of the total abundance of birds has varied a lot in the study region.\\ \\
The ultimate goal of developing this methodology is to integrate the different sources of bird count data to predict total abundance of birds across Norway, information that will be used in further studies of human impact on biodiversity, including predicting bird mortality hotspots due to powerlines and wind farms \citep{Bernardino2018,Bevanger1995,Bevanger2001,Serrano2020}. Therefore, achieving a good predictive performance of our models is of paramount importance to properly assess the vulnerability of different regions to human development based on the total local abundance of birds. Although we found differences in goodness of fit between the three models, the differences in predictive performance were small. However, a flexible model specification seemed the best choice for ensuring good predictions. For example, model 3 (which included $\omega_2(\textbf{s})$ to account for particularities of the line counts in Norway) yield the most accurate predictions at the observed locations in Norway. This is associated to the extra complexity found between line transects and point counts in Norway, which unlike the two sampling protocols in Sweden, did not have a clear linear relation, as they are only complementary to one another.\\ \\
In conclusion, in this paper we propose models to integrate multiple professional surveys with differences in their sampling protocols. These differences are usually determined by the country of origin of the data (sampling protocol) or by the specific targets of each monitoring scheme. The INLA-SPDE approach implemented in the R-INLA package makes it straightforward to perform full Bayesian inference for models that integrate multiple sources of information, even if they are not standardized or report the observed counts in different units. A natural extension of this work is the application of the proposed modeling framework to solve a broader range of ecological questions at larger geographical scales or for species with poor data \citep{buckland2017} that incorporate more sources of information given its convenience and simple implementation.

%% file: acknowledgements.tex
\section{Acknowledgements}
The Norwegian terrestrial bird monitoring (TOV-E) is coordinated by BirdLife Norway and Norwegian Institute for Nature Research, and is financed by the Ministry of Climate and Environment and the Norwegian Environment Agency. The Swedish Bird Survey is supported by grants from the Swedish Environmental Protection Agency, with additional financial and logistic support from the Regional County Boards (Länsstyrelsen).